\documentclass[11pt]{iopart}

\newcommand{\fnl}{f_{\mathrm{NL}}}

\newcommand{\MP}{M_{\rm P}}
\newcommand{\gsim}
 {\mbox{\raisebox{-1.ex}{$\stackrel{\textstyle>}{\textstyle\sim }$}}}
\renewcommand{\lsim}
 {\mbox{\raisebox{-1.ex}{$\stackrel{\textstyle<}{\textstyle\sim}$}}}

\newcommand{\XR}{Z\dot{R}^2}
\newcommand{\Vol}{\mathrm{Vol} (X_5)}
\newcommand{\PX}{P_{,X}}

\newcommand{\YAdS}{Y_{\mathrm{AdS}}}

\newcommand{\Ps}{\mathcal{P}_S^2}
\newcommand{\Pt}{\mathcal{P}_T^2}

\usepackage{setstack}
\usepackage{iopams}

\begin{document}
\title{Gravitational Wave Constraints on Multi-Brane Inflation}
\date{\today}
\author{Ian Huston and James E. Lidsey}
\address{Astronomy Unit, School of Mathematical Sciences\\
  Queen Mary, University of London\\
  Mile End Road, London E1 4NS\\
  United Kingdom}

\author{Steven Thomas}
\address{Centre for Research in String Theory, Department of Physics\\
Queen Mary, University of London, \\
Mile End Road, London E1 4NS \\
United Kingdom}

\author{John Ward}
\address{Department of Physics and Astronomy\\
University of Victoria, Victoria \\
BC, V8P 1A1, Canada}
\eads{\mailto{I.Huston@qmul.ac.uk}, \mailto{J.E.Lidsey@qmul.ac.uk},
\mailto{S.Thomas@qmul.ac.uk}, \mailto{jwa@uvic.ca}}
\begin{abstract}
A class of non-canonical inflationary models is identified, where the
leading-order contribution to the non-Gaussianity of the curvature 
perturbation is determined by the sound speed of the fluctuations
in the inflaton field.
Included in this class of models is the effective action for 
multiple coincident branes in the finite $n$ limit. The action for this
configuration is determined using a powerful iterative technique, based
upon the fundamental representation of $SU(2)$.
In principle the upper bounds on the tensor-scalar ratio that arise in the
standard, single-brane DBI inflationary scenario can be relaxed in 
such multi-brane configurations if a large and detectable non-Gaussianity 
is generated. Moreover models with a small 
number of coincident branes could generate a gravitational wave 
background that will be observable to future
experiments.

\vspace{3mm}
\begin{flushleft}
  \textbf{Keywords}:
  physics of the early universe, inflation, string theory and cosmology
\end{flushleft}

\end{abstract}
\maketitle

\section{Introduction}

The quest to realise inflation within string/M-theory continues to 
attract considerable attention. The Dirac-Born-Infeld (DBI) scenario 
of the compactified type IIB theory is a well-motivated model, 
in which inflation is driven by one or more ${\rm D}$-branes 
propagating in a warped `throat' background
\cite{brane1,brane11,brane12,brane13,brane2,brane20,brane3,brane4,brane5,brane6}
. Such a background is generated 
by the non-trivial form-field fluxes over the internal dimensions. 
(For recent reviews, see
\cite{tyereview,McAllister:2007bg,Lorenz:2007ze,Kallosh:2007wm,Bean:2007eh,bean,
cline}.) 
In the simplest version of the scenario, 
the inflaton parametrizes the radial 
position in the throat of a single ${\rm D3}$-brane. 
The brane dynamics are determined by the DBI action in such a 
way that the inflaton's kinetic energy is bounded from above by the warped 
brane tension. The regime where this bound is nearly saturated is 
known as the `relativistic' limit.

Recently relativistic DBI inflation has come under considerable 
pressure when confronted with cosmological observations.  
Baumann \& McAllister (BM) and Lidsey \& Huston (LH) 
have shown that the ratio of the amplitudes of the 
tensor and scalar perturbations generated during 
inflation is bounded from above by $r \lsim 10^{-7}$
\cite{bmpaper,lidseyhuston}. 
However in ultra-violet (UV) versions of the scenario, where 
the ${\rm D}$-brane is moving towards the tip of the throat, the tensor-scalar 
ratio is also bounded from below, $r \gsim 0.1 (1-n_s)$, 
where $n_s$ denotes the spectral index 
of the scalar perturbation spectrum \cite{lidseyhuston}. 
The two bounds on $r$ are incompatible 
if $n_s \sim 0.95$, as currently favoured by Cosmic Microwave Background 
(CMB) observations \cite{spergel,Komatsu:2008hk}. 

The purpose of the present paper is to investigate whether the 
upper bounds on $r$ can be relaxed in more general DBI 
inflationary scenarios. 
A natural extension to the single ${\rm D3}$-brane model 
is to consider a ${\rm D}p$-brane wrapped around a $(p-3)$-cycle of the 
internal space. For example, Becker \etal \cite{Becker:2007ui} 
have proposed a model where inflation is driven by a ${\rm D5}$-brane. 
In this case, the range of allowed values for the inflaton 
becomes independent of the throat charge, $N$, which weakens the upper bound on 
the tensor-scalar ratio to $r \lsim 0.04$. Strictly speaking this is 
only true for the ${\rm D7}$-brane case, since the
wrapped ${\rm D5}$-brane imposes 
$\Delta \phi \sim N^{-1/4}$.
However in arriving at this bound, it was assumed that the 
backreaction effects of any fluxes in the throat were 
negligible. Kobayashi \etal \cite{Kobayashi:2007hm} 
considered both ${\rm D5}$- and
${\rm D7}$-brane models, but concluded that the former case
required an excessively large background charge in 
order to relax the bounds on $r$. Whilst this is highly constraining, it is still much better than the
case for single ${\rm D3}$-branes which require a much higher background charge - and therefore are effectively ruled
out as a predictive model. Thus wrapped brane configurations are preferable to single brane models. However
the difficulty in these models is that the backreaction is no longer under control.

Alternative ways to relax these bounds have been proposed, including
theories based upon multi-field models \cite{Huang:2007hh}, the addition of
angular momentum as another degree of freedom \cite{spinflation} and using
different throat geometries \cite{Gmeiner:2007uw}. However it must be noted that
the extra degrees of freedom introduced in these models do not solve the problem. The 
bounds are relaxed only by a small fraction, and therefore these models should still be 
regarded as being unsatisfactory since they require an extreme amount of fine tuning in order
to work.

Another alternative possibility is to consider 
multiple brane configurations\footnote{In certain limits this approach 
is actually dual to considering wrapped branes \cite{Ward:2007gs}.}. 
In the case where  
$n$ branes are localised initially at equal distances $l > l_s$ and 
subsequently follow the same trajectory, 
the effective theory is equivalent to that of $n$ copies of the
action for a single brane. A more general initial condition, particularly
for branes created in the infra-red (IR) region of the throat
\cite{brane13, DeWolfe:2004qx, Kachru:2002gs}, is that
the branes should be separated over a range of scales, 
with a subset being coincident and the remainder being widely separated. 

Our approach in this paper is two-fold. We 
begin in Sections \ref{sec:noncanoninfl} and \ref{sec:theobound} 
by noting that the upper bounds on the tensor-scalar ratio arise due to the
special algebraic properties of the DBI action. We 
then adopt a phenomenological approach in Section \ref{sec:relaxingbound}
and identify a general class of non-canonical inflationary models, 
where the leading-order contribution to the non-Gaussianity of the 
curvature perturbation is determined 
entirely by the speed of sound of the inflaton fluctuations. 
In these models, the bounds on $r$ can be relaxed 
if significant non-Gaussianities are generated.  
This class of models includes the relativistic limit  
of the action for $n$ coincident ${\rm D3}$-branes, 
which originates from a UV complete theory. This motivates us 
to develop the theory of multiple coincident branes further in Section
\ref{sec:multibranes}. We find that for a finite number of branes, 
the effective action for $n$ coincident branes can be derived and the backreaction
kept firmly under control.
In Section \ref{sec:multibounds} we find that such models 
can in principle lead to a detectable 
gravitational wave background if 
the number of coincident branes is sufficiently small. 

Units are chosen such that $\hbar = c =1$ and $\MP \equiv (8\pi G)^{-1/2}
= 2.4 \times 10^{-18} \, {\rm GeV}$ denotes the 
reduced Planck mass. 

\section{Non-Canonical Inflation} \label{sec:noncanoninfl}

The low-energy, world-volume dynamics of a 
${\rm D3}$-brane in a warped background is determined 
by an effective action of the form 
\begin{equation}
\label{DBIaction}
S=\int  d^4x \sqrt{|g|} \left[ \frac{\MP^2}{2} R 
+ P (\phi , X) \right] \,,
\end{equation}
where $R$ is the Ricci curvature scalar, 
$X \equiv -\frac{1}{2}g^{\mu \nu}\nabla_\mu \phi \nabla_\nu \phi$
denotes the kinetic energy of the inflaton field $\phi$, and the function  
$P (\phi , X)$ is referred to as the `kinetic function'.  

We assume that the four-dimensional universe is   
spatially flat and isotropic and sourced by an  
homogeneous inflaton field, $\phi =\phi (t)$, with energy 
density $E = 2X\PX - P$, where a subscripted comma denotes partial
differentiation. 
We further assume that the inflaton dynamics  
generates a quasi-exponential expansion of the universe, 
where $\epsilon \equiv -\dot{H}/H^2 \ll1$. 
 
It proves convenient to define two parameters in terms of the 
kinetic  function $P$ and its derivatives \cite{lidser1,lidser3}: 
\begin{eqnarray}
\label{defcs}
 c_s^2 \equiv \frac{\PX}{\PX + 2X P_{,XX}} \,,
\\
\label{deflambda}
\Lambda \equiv  \frac{X^2 P_{,XX} +
\frac{2}{3}X^3 P_{,XXX}}{X P_{,X} +
2X^2 P_{,XX}}\,.
\end{eqnarray}
The first parameter, $c_s$, determines the sound speed of fluctuations 
in the inflaton field. This can be significantly less than unity, 
in contrast to slow-roll inflation driven by a canonical 
field such that $P_{,X} =1$.

The amplitudes of the scalar and tensor perturbations 
generated during inflation are given by \cite{gm}
\begin{eqnarray} 
\label{eqn:Ps2}
 \Ps = \frac{H^4}{8\pi^2 X}\frac{1}{c_s \PX} \,,
\\
\label{eqn:Pt2}
\Pt = \frac{2}{\pi^2} \frac{H^2}{\MP^2} \,,
\end{eqnarray}
respectively, and the ratio of these amplitudes 
is defined as \cite{gm} 
\begin{equation}
\label{defr}
r\equiv \frac{\Pt}{\Ps} = 16c_s \epsilon \,.
\end{equation}  
The WMAP3 normalization of the CMB power spectrum 
implies that $\Ps= 2.5\times10^{-9}$ and 
the experimental upper bound on the tensor-scalar 
ratio is $r <0.55$ \cite{spergel}.

Deviations from Gaussian statistics in the curvature perturbation, ${\cal{R}}$,
are parametrized in terms of the non-linearity parameter, 
$\fnl$, which is defined by ${\cal{R}} = {\cal{R}}_G + \frac{3}{5} \fnl  (
{\cal{R}}_G^2 -\langle {\cal{R}}_G^2 \rangle )$, where the 
quadratic component represents a convolution and 
${\cal{R}}_G$ denotes the Gaussian contribution \cite{maldacena}. In the limit  
where the three momenta have equal magnitude (corresponding to the equilateral  
triangle limit), the leading-order contribution to the non-linearity 
parameter is given by \cite{chenetal,lidser3}
\begin{equation} 
\label{deffnl}
 \fnl = -\frac{35}{108}\left(\frac{1}{c_s^2} -1 \right) +
\frac{5}{81}\left( \frac{1}{c_s^2} -1 -2\Lambda \right) \,.
 \end{equation} 
One should note that the sign convention is that employed
by the WMAP data set.
Data from WMAP3 imposes the bound $|\fnl| < 300$ on this parameter
\cite{spergel}. The corresponding bounds for other triangle configurations 
may be much tighter than this and this may be particularly relevant if 
non-Gaussian signatures have indeed been detected in the 
CMB \cite{Yadav:2007yy,crim}. The more recent WMAP5 data set
\cite{Komatsu:2008hk} improves on this bound somewhat, and
also indicates that it is distinctly asymmetric. At the $95 \%$ confidence level, the bound on the 
equilateral triangle becomes $-151 < \fnl < 253$.

Eqs. (\ref{eqn:Ps2}) and (\ref{eqn:Pt2}) imply 
that the variation of the inflaton field during inflation  
is related to the tensor-scalar amplitude by \cite{lyth,bmpaper}
\begin{equation}
\label{genlythbound}
\frac{1}{\MP^2}
\left( \frac{d \phi}{d \cal{N}} \right)^2 = \frac{r}{8 c_s P_{,X}} \,,
\end{equation}
where ${\cal{N}} \equiv \int dt \, H$ denotes the number of e-foldings.
We will refer to the epoch of inflation that can be directly 
constrained by cosmological observations as 
`observable inflation' and will assume that this phase 
occurred when the brane was located within a 
throat region\footnote{We denote the values of all parameters 
evaluated during observable inflation by a subscript 
`$*$'.}. Observable inflation corresponds to no more than about 4 e-foldings  
of inflationary expansion, $\Delta {\cal{N}}_* \simeq 4$. 
The total variation in the inflaton field between the epoch of observable 
inflation and the end of inflation is then given by
\begin{equation}
\label{totalfield}
\frac{\Delta \phi_{\rm inf}}{\MP} = 
\left( \frac{r}{8c_sP_{,X}} \right)_*^{1/2} {\cal{N}}_{\rm eff} \,,
\end{equation}
where
\begin{equation}
\label{Neff}
{\cal{N}}_{\rm eff} \equiv \left( \frac{c_sP_{,X}}{r}\right)_*^{1/2}
\int_0^{{\cal{N}}_{\rm end}}  
\left( \frac{r}{c_sP_{,X}} \right)^{1/2} d {\cal{N}} \,.
\end{equation}
If $r/(c_s P_{,X})$ varies 
sufficiently slowly during observable inflation, 
the corresponding change in the value of the inflaton  
field is given approximately by \cite{lyth,bmpaper}
\begin{equation}
\label{approxlyth}
\left( \frac{\Delta \phi}{\MP} \right)_*^2 \simeq 
\frac{(\Delta {\cal{N}}_*)^2}{8} \left( \frac{r}{c_sP_{,X}} \right)_* \,.
\end{equation}

\section{Theoretic Upper Bounds on the Tensor-Scalar Ratio}
\label{sec:theobound}

The ten-dimensional metric of the warped deformed conifold inside a  
throat region has the form 
\begin{equation}
\label{conemetric}
ds_{10}^2= h^2 ( \rho) ds_4^2 + h^{-2} (\rho ) 
\left( d\rho^2 +\rho^2 ds_{X_5}^2 \right) \,,
\end{equation} 
where the `warp factor' $h(\rho )$ is a function of the radial 
coordinate $\rho$ along the throat and $X_5$ denotes a five-dimensional, 
Sasaki-Einstein manifold. In many scenarios, the ten-dimensional 
manifold (\ref{conemetric}) can be approximated by the product 
$AdS_5 \times X_5$, where $AdS_5$ represents five-dimensional, 
anti-de Sitter space. In this case, the warp factor is given by  
\begin{equation}
h = \frac{\rho}{L} , \qquad 
L^4 = \frac{4\pi^4 g_s N}{\Vol m_s^4}\,,
\end{equation} 
where $L$ denotes the $AdS_5$ radius of curvature, 
$\Vol$ is the volume of the five-manifold $X_5$ with unit radius, 
$N$ is the ${\rm D3}$-brane charge in the throat, 
$g_s$ is the string coupling and $m_s$ is the string mass scale. 
Generally the value of the inflaton field 
is determined by the radial position of the 
${\rm D3}$-brane in the throat, 
$\phi \equiv \rho \sqrt{T_3}$, where $T_3 \equiv  
m_s^4/[(2\pi)^3g_s]$ is the brane tension. 

The four-dimensional Planck mass is related to the volume 
of the compactified Calabi-Yau three-fold, $V_6$, such that
$\MP^2 =V_6 \kappa_{10}^{-2}$, where $\kappa_{10}^2 \equiv \frac{1}{2}
(2 \pi )^7 g_s^2/ m_s^{8} = \pi /T_3^{2}$ for a 
${\rm D3}$-brane \footnote{We parameterise the Planck scale 
in terms of the ${\rm D3}$-brane tension out of convenience, 
and note that there is no physical relationship between the two.}. 
Hence the Planck mass is bounded from 
below by the volume of a throat region, 
$\MP^2 >V_{6,{\rm th}} \kappa_{10}^{-2}$, 
where $V_{6,{\rm th}} \lsim V_6$ denotes the throat volume. 
For an $AdS_5 \times X_5$ throat, Baumann \& McAllister (BM)
exploited this inequality to derive an upper limit on the 
maximum variation of the inflaton field in the throat, 
$\Delta \phi_{\rm max} < 2\MP/\sqrt{N}$, which leads to the 
corresponding limit $|\Delta \phi |_* < 2\MP /\sqrt{N}$ \cite{bmpaper}.  
Combining this with the constraint (\ref{approxlyth}) therefore 
yields an upper limit on the tensor-scalar ratio: 
\begin{equation}
\label{BMbound}
r_* < \frac{32}{N {\cal{N}}_{\rm eff}^2} \left( 
c_s P_{,X} \right)_* \,.
\end{equation}

Two of the authors (Lidsey \& Huston, LH) derived a complementary bound on the 
tensor-scalar ratio by noting that during observable inflation
the brane spans a fraction of the throat volume \cite{lidseyhuston}
\begin{equation}
\label{approxvol}
| \Delta V_{6,*} | \simeq \Vol \frac{| \Delta \rho_*| \rho_*^5}{h_*^4}
\end{equation}
and, since $| \Delta V_{6,*} |< 
V_{6,{\rm th}}$, it follows that 
\begin{equation}
\label{firstbound}
\left(\frac{\Delta \phi}{\MP}\right)^2_* < \frac{T_3 \kappa_{10}^2
(\Delta \rho_*)^2}{| \Delta V_{6*}|} \,.
\end{equation}
It was then assumed that the fractional change in the value of the 
inflaton field during observable inflation was less than unity:
\begin{equation}
\label{delta1constraint}
|\Delta \phi_* | <  \phi_* \,.
\end{equation} 
This condition is necessarily satisfied in UV
versions of the scenario, 
where the brane is moving towards the tip of the throat, but must be 
assumed as a further constraint in IR versions 
where the brane moves out of the throat, since in these latter cases
$\phi_*$ could be very small. Combining the limit 
(\ref{delta1constraint}) with Eq. (\ref{approxvol}) then implies that 
\begin{equation}
| \Delta V_{6,*} | > \Vol \frac{( \Delta \rho_* )^6}{h_*^4}
\end{equation}
and substituting this constraint into the bound (\ref{firstbound}) yields 
the condition 
\begin{equation}
\left( \frac{\Delta \phi}{\MP} \right)_*^6 < \frac{\pi T_3}{\Vol} 
\left( \frac{h_*}{\MP} \right)^4 \,.
\end{equation}
Finally substituting the constraint (\ref{approxlyth}) 
yields the upper limit \cite{lidseyhuston}
\begin{equation}
\label{LHbound}
r_* < \frac{10}{(\Delta {\cal{N}})_*^2} \left( \frac{T_3}{\Vol} \right)^{1/3} 
\left( \frac{h_*}{\MP} \right)^{4/3} \left( c_s P_{,X} \right)_* \,.
\end{equation}

Comparison of the limits (\ref{BMbound}) and (\ref{LHbound})
implies that the LH bound is the stronger of the two when
\begin{equation}
\label{LHstronger}
h_*^{4/3}N < 20 \left( \Vol g_s \right)^{1/3}  
\left( \frac{m_s}{\MP} \right)^{-4/3} 
\frac{( \Delta {\cal{N}} )_*^2}{{\cal{N}}^2_{\rm eff}} \,.
\end{equation}
For typical field-theoretic values $\Vol \simeq \mathcal{O}(\pi^3)$, $m_s \sim
0.1 \MP$ 
and  $g_s \sim 10^{-2}$, this implies 
\begin{equation}
\label{LHstronger1}
h_*^{4/3} N < 300 \frac{(\Delta {\cal{N}})^2_*}{{\cal{N}}_{\rm eff}^2} \,.
\end{equation}

In the following Section, we identify a class of models in which  
these bounds could be relaxed. 

\section{Relaxing the Upper Bounds on the Tensor-Scalar Ratio}
\label{sec:relaxingbound}

In the standard DBI scenario, the kinetic function defined in 
Eq. (\ref{DBIaction})  takes the form 
\begin{equation}
\label{DBIkinetic}
P (\phi , X) = -T(\phi ) \sqrt{1-2T^{-1}(\phi ) X} + T(\phi ) - V(\phi ) \,,
\end{equation}
where $T(\phi ) = T_3 h^4 (\phi )$ 
is the warped brane tension and  $V(\phi )$ is the 
inflaton potential. 
Typically in warped compactifications of 
IIB supergravity, this potential is determined by the 
relevant fluxes and brane interaction terms. 
We will ignore the precise origin
and form of this potential, but simply
note that it is highly sensitive to the string theoretic construction. For the
purpose of this note we will simply treat it 
as an arbitrary function of the inflaton field.
(See, for example, Ref. \cite{brane5} for a discussion 
on the precise form that the inflaton potential may take.)

The standard DBI scenario (\ref{DBIkinetic}) is algebraically special, 
in the sense that the kinetic function satisfies the constraints 
\begin{equation}
\label{special}
c_s P_{,X} = 1 , \qquad  \Lambda = \frac{1}{2} \left( 
\frac{1}{c_s^2} -1 \right) \,.
\end{equation}
It follows that the 
bounds (\ref{BMbound}) and (\ref{LHbound}) 
on the tensor-scalar ratio could in principle be 
significantly relaxed in models where $(c_sP_{,X})_* \gg 1$. 
In view of the second relation in Eq. (\ref{special}), it is of interest 
to take a phenomenological approach and consider the more  
general class of models where 
\begin{equation}
\label{defalpha}
\frac{1}{c_s^2} -1 = \alpha \Lambda \,,
\end{equation}
for some positive constant $\alpha$. Moreover since 
a large non-Gaussian signature in the curvature perturbation is 
typically generated in models where the sound speed of fluctuations 
is small, we will begin by considering 
scenarios where the kinetic function satisfies the 
inequalities: 
\begin{equation}
\label{Plimits}
X^2 P_{,XXX} \gg XP_{,XX} \gg P_{,X} \,.
\end{equation}

In these limits the constraint (\ref{defalpha}) reduces to the 
third-order, non-linear, partial differential equation
\begin{equation}
\label{pde}
P^2_{,XX} = \frac{\alpha}{6} P_{,X} P_{,XXX} \,.
\end{equation}
Changing the dependent variable to $Q \equiv P_{,XX}/P_{,X}$ 
reduces Eq. (\ref{pde}) to 
\begin{equation}
\label{Qequation}
\alpha Q_{,X} = (6-\alpha )Q^2 \,,
\end{equation}
and it is straightforward to integrate Eq. (\ref{Qequation}) 
exactly. The remaining integrations can also be performed analytically 
and the general solution to Eq. (\ref{pde}) for $\alpha \ne
6$ is given by\footnote{The special case $\alpha =6$ results in an 
exponential dependence of the kinetic function on $X$, and is 
therefore an example of a higher-derivative theory. However we 
do not consider this model further, since it does not lead to a 
weakening of the gravitational wave constraints.}
\begin{equation}
\label{gensol}
P (\phi , X) = f_1 (\phi ) \left[ 1-f_2 (\phi ) X 
\right]^m -f_3(\phi ) \,,
\end{equation}
where $f_i (\phi )$ are arbitrary functions of the scalar 
field and 
\begin{equation}
m \equiv \frac{2(\alpha -3 )}{\alpha -6} \, . 
\end{equation}
It can be verified that the inequalities (\ref{Plimits}) 
are satisfied in the `relativistic' limit, where $X \simeq 1/f_2$.
We consider the inflationary dynamics in this limit in what follows. 
For completeness, we note that Eq. (\ref{defalpha}) can be solved 
analytically in 
full generality and the solution is presented in the Appendix. 

The standard DBI scenario is recovered for $m=1/2$. 
More generally however, Eq. (\ref{gensol}) implies that  
\begin{eqnarray}
\label{consequence1}
c_sP_{,X} \simeq -\frac{m f_1 f_2}{\sqrt{2(1-m)}} \left( 
1- f_2X \right)^{(2m-1)/2} \,,
\\
c_s^2 \simeq \frac{1-f_2X}{2(1-m)} \,,
\end{eqnarray}
when $X \simeq 1/f_2$. Self-consistency therefore
requires $m<1$. Moreover
we find from Eq. (\ref{deffnl}) that 
\begin{eqnarray}
\label{consequence3}
\fnl \simeq \frac{-\beta}{1-f_2X} , \qquad \beta \equiv \frac{5(59-55m)}{486}
\,,
\\
\label{consequence4}
\fnl \simeq -\frac{\sigma}{c_s^2}, \qquad \sigma \equiv 
\frac{5}{972} \left( \frac{59-55m}{1-m} \right) \,.
\end{eqnarray}
Hence substituting Eqs. (\ref{consequence1}) and (\ref{consequence3}) 
into the BM bound (\ref{BMbound}) and the LH bound (\ref{LHbound}) implies that 
\begin{equation}
\label{weakBMbound}
r_*< \frac{32}{N {\cal{N}}_{\rm eff}^2} \frac{(-m) f_1 f_2}{\sqrt{2(1-m)}}
\left( -\frac{\fnl}{\beta} \right)^{(1-2m)/2}
\end{equation}
and 
\begin{equation}
\label{weakLHbound}
r_* < \frac{10}{(\Delta {\cal{N}})_*^2} \left( \frac{T_3}{\Vol} \right)^{1/3} 
\left( \frac{h_*}{\MP} \right)^{4/3}
\frac{(-m) f_1 f_2}{\sqrt{2(1-m)}}
\left( -\frac{\fnl}{\beta} \right)^{(1-2m)/2}
\end{equation}
respectively. 

We conclude, therefore, that 
the upper limit on the tensor-scalar ratio could be significantly 
relaxed if $m <1/2$, since the non-linearity parameter is at present only 
weakly constrained at $\fnl >-151$. Although it is possible 
to phenomenologically construct a model which has a value of $m$ in this 
range, it is clearly preferable to identify  UV complete models
that satisfy this requirement within a string theory context. 
Unfortunately this is quite difficult to achieve since the inflaton 
will either be associated with an open or closed string mode. 
The open strings are governed by relativistic actions of the 
DBI form, whilst closed strings arise from compactification of Einstein gravity
and are typically put into canonical form.
However there do exist classes of open string
models which satisfy the above requirement, 
namely those associated with multiple coincident branes.

More specifically, if the branes are
spatially separated, the effective action is algebraically equivalent 
to that of a single brane. It will therefore not satisfy the 
bound on $m$ \footnote{In this discussion, we are ignoring 
the non-trivial backreaction of these branes on the background, and therefore 
one should be careful about the range of validity of the effective action.}. 
Similarly it was shown in \cite{thomasward}
that $n$ coincident branes, in the large $n$ limit, will also fall into 
this class of models. On the other hand, if it is assumed that 
$n$ is finite, the special properties associated with the 
matrix degrees of freedom become important and this 
results in a kinetic function satisfying $m \le 1/2$.
We will discuss this in more detail in the following Section.

\section{Action for Multiple Coincident Branes} \label{sec:multibranes}

We have seen how the form of the kinetic function
$P$ can significantly change the
strength of the LH bound on the tensor-scalar ratio, depending on its explicit
form. One model in which a suitable form for 
$P$ is realised is the multiple coincident
brane model as outlined by Thomas \& Ward \cite{thomasward}.

The world-volume theory for coincident branes is not fully known, although
a number of proposals have been made. We will restrict our analysis to Myers' 
prescription, since this has been extensively discussed in the 
literature\footnote{There is also a proposal by Tseytlin \cite{Tseytlin}
for the non-Abelian theory of coincident D-branes.} \cite{myers1, myers2}. 
In general the open string degrees of freedom for $n$ coincident branes 
combine to fill out representations of $U(n)$ (as opposed to $U(1)^n$ 
in the case of separated branes). This introduces a non-Abelian 
structure into the theory. In the single brane case, the fluctuations of the 
brane are characterised by induced scalar fields on the world-volume. 
However for multiple branes
these scalars must be promoted to matrix representations of some gauge group. 

Typically the transverse space of any given compactification will always admit
an $SO(3)$ isometry. We can therefore choose our scalars to 
transform under representations of the algebra of $SO(3) \sim SU(2)$ by making 
the identifications
\begin{equation}
\phi^i = R \alpha^i \hspace{1cm} i =1,  \ldots 3 \,,
\end{equation}
where $R$ is some scale with canonical mass dimension, and the $\alpha^i$ are
specified to be the irreducible generators satisfying the commutator
\begin{equation}
[\alpha^i, \alpha^j] = 2i \epsilon_{ijk} \alpha^k \,,
\end{equation}
and the conditions
\begin{equation}
\frac{1}{n} Tr(\alpha^i \alpha^j) = \hat{C} \delta^{ij} = (n^2-1) \delta^{ij}
\,,
\end{equation}
where $\hat{C}$ is the quadratic Casimir of the gauge group.
The irreducibility condition corresponds to the configuration being in the
lowest energy state. It is therefore an additional fine-tuning
of the initial conditions. 

The Myers prescription requires a symmetrised trace (denoted $STr$) to 
be made over the gauge group. This implies that 
the symmetric averaging must be taken over all the group dependence 
before taking the trace. For $n \gg 1$, the symmetric trace can be approximated with a trace, 
which results in the usual DBI action multiplied by a potential term (as
described in \cite{thomasward, Kachru:2002gs}). 
However for finite $n$, the symmetrisation clearly becomes more important and it is essential that we
have some means of performing this operation. Recently
a prescription for the symmetric trace at finite $n$ was proposed \cite{Ramgoolam:2004gw, McNamara:2005ry}, 
using highest weight methods and chord diagrams. 

The result is that the $STr$ acts on different spin representations of $SU(2)$ 
in the following manner:
\begin{eqnarray}
STr (\alpha^i \alpha^i)^q = 2(2q+1)\sum_{i=1}^{n/2}(2i-1)^{2q} , 
\qquad n\; \mathrm{ even}\,, 
\\
STr (\alpha^i \alpha^i)^q = 2(2q+1)\sum_{i=1}^{(n-1)/2} (2i)^{2q} , 
\qquad n\; \mathrm{ odd}\,.
\end{eqnarray}

In order for the solution to converge in this prescription, 
it is also necessary to modify the definition of the radius of the 
$SU(2)$ sphere. In the large $n$ limit, this is given by
\begin{equation}
\rho^2 = \lambda^2 R^2 \frac{1}{n} Tr(\alpha^i \alpha^i) = \lambda^2 R^2
\hat{C} \,,
\end{equation}
where $\lambda \equiv 2\pi l_s^2 = 2\pi m_s^{-2}$, 
whereas for finite $n$, it becomes
\begin{equation}
\rho^2 = \lambda^2 R^2 \mathrm{Lim}_{q \to \infty} \left(\frac{STr (\alpha^i
\alpha^i)^{q+1}}{STr(\alpha^i \alpha^i)^q} \right) 
= \lambda^2 R^2 (n-1)^2 \,.
\end{equation}
This converges to the large $n$ result in the appropriate limit.
This point is important, since the warp factor 
of the four-dimensional theory is typically of the form $h= h(\rho)$.

The resulting kinetic function for $n$ coincident branes in 
the finite $n$ limit is therefore given by
\begin{equation}
\label{generalP}
\fl
P = -T_3 STr \left(h^4(\rho) \sum_{k,p=0}^{\infty}(-\XR)^k Y^p (\alpha^i
\alpha^i)^{k+p}{1/2 \choose k} {1/2 \choose p} + V(\rho)
-
h^4(\rho) \right) \,,
\end{equation}
where 
\begin{equation}
\fl
Z \equiv \lambda^2 h^{-4}(\rho), \hspace{0.5cm} Y \equiv 4\lambda^2 R^4
h^{-4}(\rho),
\hspace{0.5cm} {1/2 \choose q}
\equiv \frac{\Gamma(3/2)}{\Gamma(3/2-q)\Gamma(1+q)} \,.
\end{equation}
Note that the second and third terms in Eq. (\ref{generalP}) 
are singlets under the $STr$ and therefore contribute terms proportional 
to $n$. The physics of these branes away from the large $n$ limit is particularly interesting as discussed further
in \cite{thomasward, Ward:2007gs}.

The simplest case to consider is that of two coincident branes. However
the form of the $STr$ prescription implies that all other solutions for 
$n>2$ can be deduced entirely from the $n=2$ solution by a recursion relation. 
In order to see this, let us define 
\begin{eqnarray}
\label{2brane}
P_2(Z,Y) &=& - 2 T_3 h^4 \left(\frac{(1+2Y -
(2+3Y)\XR)}{\sqrt{1+Y}\sqrt{1-\XR}}\right) \,, \nonumber \\
E_2(Z,Y) &=& 2 T_3 h^4 \left(\frac{(1+2Y -
Y\XR)}{\sqrt{1+Y}(1-\XR)^{3/2}}\right) \,.
\end{eqnarray}
These quantities correspond to the pressure and energy density functions 
when $n=2$ which arise solely from the DBI sector of the action.
The full pressure and energy densities are then 
given by $P = P_2 - 2T_3(V-h^4)$ and
$E = E_2 + 2T_3 (V-h^4)$, respectively.
Since the symmetrised trace acts differently on the differing spin
representations of $SU(2)$, we should expect this structure to follow
through in the recursion relation. Indeed, we find that for odd $n$
\begin{eqnarray}
\label{oddbrane}
P_n^{(O)} &=& \left(\sum_{k=1}^{(n-1)/2} P_2[(2k)^2Z, (2k)^2Y]
\right)-nT_3(V-h^4) \,, \nonumber \\
E_n^{(O)} &=& \left(\sum_{k=1}^{(n-1)/2} E_2[(2k)^2Z, (2k)^2Y] \right)+
nT_3(V-h^4) \,,
\end{eqnarray}
whilst for even $n$ we find that 
\begin{eqnarray}
\label{evenbrane}
P_n^{(E)} &=& \left(\sum_{k=1}^{n/2} P_2[(2k-1)^2Z, (2k-1)^2Y]
\right)-nT_3(V-h^4) \,, \nonumber \\
E_n^{(E)} &=& \left(\sum_{k=1}^{n/2} E_2[(2k-1)^2Z, (2k-1)^2Y] \right)+
nT_3(V-h^4) \,.
\end{eqnarray}

For example, we can employ these recursion relations to obtain the 
solutions for $n=3$:  
\begin{eqnarray}
P &=& - 2T_3 \left(\frac{h^4
(1+8Y-8\XR(1+6Y))}{\sqrt{1-4\XR}\sqrt{1+4Y}}\right)
 -3T_3(V-h^4) \,, \nonumber \\
E &=&  2T_3 \left(\frac{h^4
(1+8Y(1-2\XR))}{(1-4\XR)^{3/2}\sqrt{1+4Y}}\right) +3T_3(V-h^4) \,,
\end{eqnarray}
which agrees precisely with the result computed by direct 
expansion of the $STr$ prescription. Furthermore the $n=4$ case is given by 
\begin{eqnarray}
\fl
P = -2T_3\left(\frac{h^4
(1+2Y-\XR(2+3Y))}{\sqrt{1+Y}\sqrt{1-\XR}} \right. 
\nonumber \\
\left. +\frac{h^4(1+18Y-9\XR(2+27Y))}
{\sqrt{1+9Y}\sqrt{1-9\XR}}\right) 
 -4T_3(V-h^4) \,, \nonumber \\
\fl
E =  2T_3 \left(\frac{h^4(1+2Y-Y\XR)}{\sqrt{1+Y}(1-\XR)^{3/2}} +
\frac{h^4(1+18Y-81\XR)}{\sqrt{1+9Y}(1-9X\dot{R}^2)^{3/2}} \right)
+4T_3(V-h^4) \, .
\end{eqnarray}
It it clear that the relevant functions increase in complexity as 
$n$ increases, since there are progressively more
terms to include in the $STr$ expansion. However Eqs. (\ref{oddbrane}) and (\ref{evenbrane}) represent the most general
solutions. 

One should also be aware that the backreaction of multiple branes will
typically introduce corrections of the form $n/N$, therefore it is important
for this ratio to be small in order for us to trust the supergravity analysis.
Typically we can argue that wrapped branes are dual to multi-brane configurations when
we are in the limit that $n>>1$. However since we also wish to keep $N>>1$ we must tune
the solution so that $n/N<<1$ is satisfied. Therefore the origin of the backreaction effects
is much clearer from this perspective. One can compute the $1/N$ corrections to the multi-brane
action in the large $n$ limit \cite{Ward:2007gs} which, in the dual picture, correspond to backreactive corrections
to the wrapped brane models. It would certainly be more useful to develop both these models in more detail.

\section{Bounds on the Tensor-Scalar Ratio for Multi-Brane 
Inflation} \label{sec:multibounds}

The last terms appearing in the summations of Eqs. (\ref{oddbrane}) 
and (\ref{evenbrane}) correspond  to the $k=(n-1)/2$ 
term when $n$ is odd and to the $k=n/2$ term when $n$ is even. This 
implies that for all $n$, these terms can be expressed in the form 
\begin{equation}
\label{unifiedgenP}
\fl
P = -2T_3 \left\{ \frac{h^4 \left[ 1+2(n-1)^2Y
- [ 2+3(n-1)^2Y] (n-1)^2\XR  \right]}{\sqrt{1+(n-1)^2Y}
\sqrt{1-(n-1)^2\XR}} 
 \right\} -n T_3 \left( V -h^4 \right) \,.
\end{equation}

Inspection of Eqs. (\ref{2brane})-(\ref{evenbrane}) implies that 
the relativistic limit is realised for any finite number of branes when 
$(n-1)^2 \XR \rightarrow 1$. In this case, the dominant contribution 
to the summations appearing in Eqs. (\ref{oddbrane}) and (\ref{evenbrane})
will arise from the last term, Eq. (\ref{unifiedgenP}). In this limit, 
therefore, the kinetic function appearing in the effective action simplifies to 
\begin{equation}
\label{unifiedP}
P = 2T_3 \left\{ h^4 \sqrt{1+(n-1)^2Y} 
\left( 1- \frac{2X}{T_3h^4} \right)^{-1/2} 
 \right\} - n T_3 \left(V - h^4 \right) \,,
\end{equation}
where 
\begin{equation}
\label{defY}
Y \equiv \frac{4}{(n-1)^4 \lambda^2 T_3^2} \left( \frac{\phi}{h} \right)^4 \,,
\end{equation}
\begin{equation}\label{defXR2}
\XR \equiv \frac{2}{(n-1)^2 h^4 T_3}X \,,
\end{equation} 
and we have effectively imposed the relativistic condition 
\begin{equation}
\label{relalimit}
X \simeq \frac{1}{2} T_3 h^4 \,,
\end{equation}
in the numerator of Eq. (\ref{unifiedgenP}).  
For the $n=2$ and $n=3$ cases, we have verified by direct 
calculation that when one calculates the speed of sound 
(\ref{defcs}) and the non-linearity parameter 
(\ref{deffnl}) from the general expressions (\ref{2brane}) and (\ref{oddbrane}) 
and then imposes the relativistic
limit (\ref{relalimit}), one arrives at the identical result 
by starting explicitly with Eq. (\ref{unifiedP}).

At this point we should consider the validity of 
the function in Eq. (\ref{unifiedP}). Using the
recursion relations defined in the previous section, 
we see that in the large $n$ limit the kinetic function converges to 
the corresponding function defined in the large $n$ 
limit in \cite{thomasward}. This is not the same function as that for
$n$ separated branes, as the matrix degrees of freedom 
lead to an additional potential term for the scalars. However it does belong
to the same class of models with $m=1/2$. We have 
verified this convergence numerically since the algebraic sums are
unfortunately not tractable. The key point is that there must exist some 
value of $n$, beyond which the function appears to look
more like the standard DBI action, rather than the 
approximate form proposed in (\ref{unifiedP}). For a range of background 
solutions, the numerics suggest that the approximation is 
valid up to terms of $\mathcal{O}(10)$. Since there are a large number of 
parameters in the theory, it is possible to find solutions 
where $n \gg 10$. However we will then be forced to 
generate a larger background flux, which will result 
in a situation where even the conformal 
Calabi-Yau condition is no longer valid. In view of this, we focus on
the sector of the theory where $n \le 10$, which implies that the 
backreaction is under control and that the kinetic function
is still of the required form. 

Eq. (\ref{unifiedP}) is precisely of the form given by the 
general solution (\ref{gensol}), where $m=-1/2$ and 
\begin{equation}
\label{f1}
f_1 (\phi) = 2T_3 h^4 \sqrt{1+(n-1)^2Y} , \qquad 
f_2 (\phi) = \frac{2}{T_3 h^4} \,.
\end{equation}
We may therefore immediately conclude from Eq. (\ref{consequence4}) that $\fnl
\simeq -0.3/c_s^2$. Moreover, since $\beta \simeq 0.9$ in this scenario, 
Eqs. (\ref{consequence1}) and (\ref{consequence3}) reduce to  
\begin{equation}
\label{csPX}
c_sP_{,X} \simeq -1.3 \sqrt{1+(n-1)^2Y} \fnl \,.
\end{equation}

We first consider the LH bound (\ref{LHbound}). This applies at least for all
UV scenarios. It follows after substitution of the relativistic limit
(\ref{relalimit}) into the scalar perturbation amplitude, Eq. (\ref{eqn:Ps2}),
that 
\begin{equation}
\label{usefulPs}
\Ps \simeq -\frac{1}{50} \frac{H^4}{T_3 h^4\sqrt{1+(n-1)^2Y}}
\frac{1}{\fnl}\,.
\end{equation}
Substituting the tensor-scalar ratio (\ref{defr}) into  
Eq. (\ref{usefulPs}) then results in a constraint on the magnitude of 
the warp factor during observable inflation: 
\begin{equation}
\label{warpfactor}
\frac{h^4_*}{\MP^4} \simeq \frac{-1}{2 T_3 \sqrt{1+(n-1)^2Y}} 
\frac{r^2 \Ps}{\fnl} \,.
\end{equation}
Eqs. (\ref{csPX}) and (\ref{warpfactor}) may now be substituted into 
the LH bound (\ref{LHbound}) to yield 
\begin{equation}
\label{actualLHbound}
r_* < \frac{1100}{(\Delta {\cal{N}} )_*^6} 
\frac{[1+(n-1)^2Y]}{\Vol} \Ps \fnl^2 \,.
\end{equation}

It is clear that the parameter $Y$ 
must be sufficiently large if the tensor perturbations 
are to be non-negligible. For the $AdS_5 \times X_5$ throat, this parameter  
takes the constant value    
\begin{equation}
\label{eqn:YAds}
\YAdS \equiv \frac{4\pi^2 g_s N}{(n-1)^4 \Vol} \,.
\end{equation}
In what follows, we chose natural field-theoretic values for the volume, 
$\Vol \simeq \pi^3$, and the string coupling, 
$g_s \simeq 10^{-2}$, and further assume that 
$(n-1)^2 Y \gg 1$. It is possible that observations will probe a 
range of scales $\Delta {\cal{N}}_* \simeq 1$, 
but it is more realistic to require that 
the tensor-scalar ratio should not change significantly over the 
entire range of scales that are accessible to cosmological observation,  
which corresponds to $\Delta {\cal{N}}_* \simeq 4$.
After substitution of the above values, therefore, 
the bound (\ref{actualLHbound}) simplifies to 
\begin{equation}
\label{adsupperbound}
r_* < 2.8 \times 10^{-13} \frac{N}{(n-1)^2} \fnl^2 \,.
\end{equation}

Global tadpole cancellation constrains the magnitude of
the background charge $N$ in terms of the topology of 
a Calabi-Yau four-fold such that $N < \chi /
24$, where $\chi$ is the Euler characteristic of the four-fold
\cite{witten1,witten2,witten3,sethi,gkp,klemm}. 
The  maximal known value of the Euler number for such four-folds arises from 
hypersurfaces in weighted projective spaces and is given by 
$\chi = 1, 820,448$ \cite{klemm}. This implies the upper limit of 
\begin{equation}
\label{Nlimit} 
N < 75852
\end{equation}
for known solutions, although in principle higher values are possible. 
Imposing the WMAP5 bound $\fnl>-151$ in (\ref{adsupperbound})
and noting that $n \ge 2$ for consistency then implies an absolute
upper limit 
on the tensor-scalar ratio: 
\begin{equation}
\label{absupperlimit}
r_* < 5 \times 10^{-4} \, .
\end{equation}

This limit is below the sensitivity of the Planck satellite 
$(r \gsim 0.02 )$ \cite{planck}. On the other hand, 
the projected sensitivity of future CMB polarization experiments 
indicates that a background of primordial 
gravitational waves with $r_* \gsim 10^{-4}$ 
should be observable \cite{songknox,vpj}. In view of this, 
it is interesting to consider whether
a detectable gravitational wave background could in principle 
be generated in this class of multi-brane inflationary 
models. We find from (\ref{adsupperbound}) that this would require 
\begin{equation}
\label{nlimit}
n < 1 -5.3 \times 10^{-5} \sqrt{N} \fnl < 1-0.014 \fnl \,,
\end{equation}
where the theoretic limit ({\ref{Nlimit}) for 
known compactifications has been imposed in the 
second inequality. We may deduce, therefore, that  
since we require $n \ge 2$ for consistency, a detectable tensor 
signal will require $\fnl < -70$, which implies that an observation of 
the tensors should also be 
accompanied by a sufficiently large -- and detectable -- non-Gaussianity. 
In other words, this class of models could  
be ruled out if tensors are observed in the absence of any
non-Gaussianity. On the other hand, the current 
limit of  $\fnl >-151$ implies that $n \le 3$ is required 
for the tensors to be observable. 
Consequently, if tensor perturbations are detected, this would rule 
out all models with $n \ge  4$ or, alternatively, would require presently 
unknown configurations with $N$ exceeding bound (\ref{Nlimit}). 

In the above analysis we assumed that the string coupling 
took the value $g_s \simeq 10^{-2}$. For the $AdS_5 \times X_5$ throat, 
the bound (\ref{actualLHbound}) depends proportionally on $g_s$ and can 
therefore be weakened by allowing for larger values of the string coupling. 
For example, increasing this parameter by a factor of $4$ 
to $g_s \simeq 0.04$ (so that it is still in the perturbative regime)
relaxes the limit on the number of branes for the tensors to be detectable to 
$n \le 5$. Similarly, considering a smaller value for the 
volume of the Einstein manifold $X_5$ will also weaken the upper limit. 

Let us re-iterate that this limit on $n$ is well within the 
regime of validity for the theory, which we have argued is 
self-consistent for $n<10$. Moreover since the constraint (\ref{nlimit})
arises using the absolute maximal bound on the known 
Euler characteristics, it suggests that in realistic scenarios $n$ will 
always be much smaller than this. Indeed, one could argue that 
only the $n=2$ and $n=3$ theories are
likely to be valid over a large distribution of the flux landscape. 

We must also ensure that our approximation $(n-1)^2Y \gg 1$ 
is valid for consistency.  
For the parameter values we have chosen this requires that 
$g_s N \gg  (n-1)^2$ 
and this is satisfied if the condition (\ref{nlimit}) 
holds. Note also that we require $N \gg n$ for the supergravity 
approximation to be under control and for backreaction effects to 
be negligible. This is also satisfied when (\ref{nlimit})  
holds. 

For completeness we should also consider the 
BM bound (\ref{BMbound}) for this class of models. This is given by 
\begin{equation}
\label{BMads}
r_* < -\frac{42}{N {\cal{N}}^2_{\rm eff}}\sqrt{1 +(n-1)^2Y}\fnl 
\end{equation} 
and, in the case of an $AdS_5 \times X_5$ throat, simplifies to 
\begin{equation}
\label{eq:bmadsbound}
r_* < -\frac{5}{{\cal{N}}^2_{\rm eff}} 
\frac{\fnl}{(n-1)\sqrt{N}} \,.
\end{equation} 
Comparing the limits (\ref{adsupperbound}) and (\ref{eq:bmadsbound}) 
implies that the LH bound is stronger than the corresponding BM bound if  
\begin{equation}
\label{LHstrongerads}
n > 1 -5.5 \times 10^{-14} N^{3/2} {\cal{N}}_{\rm eff}^2 \fnl 
\end{equation}
and this condition is always satisfied if 
\begin{equation}
\label{always}
-5.5 \times 10^{-14} N^{3/2} {\cal{N}}_{\rm eff}^2 \fnl  <1  \, .
\end{equation}
Moreover, the bound (\ref{always}) will itself be satisfied for 
all values of $\fnl$ and $N$ if it is satisfied when the limits 
$\fnl =-151$ and $N=75852$ are imposed. Hence, we conclude that the LH bound 
is stronger for ${\cal{N}_{\rm eff}} < 75$. 
In general, it is difficult to quantify 
the magnitude of $\mathcal{N}_{\mathrm{eff}}$ without 
imposing further restrictions on the parameters of the models 
and, in particular, on the functional form of the inflaton potential. 
However, if the ratio $\epsilon/P_{,X}$ remains approximately 
constant during the final stages of inflation, one would anticipate that 
${\cal{N}}_{\rm eff} \lsim 60$. Nevertheless, if $N \ll 75852$, the bound 
(\ref{LHstrongerads}) will only be violated for $n \le 3$ if 
${\cal{N}}_{\rm eff} \gg 60$.  

Finally, it should be emphasized that the derivation of the LH bound 
underestimates the Planck mass by assuming that 
the volume of the throat is much smaller 
than the volume of the compactified Calabi-Yau 
three-fold. It is likely, therefore, 
that the actual constraint on $r$ would be much stronger. Consequently, 
although the bound (\ref{nlimit})  
does marginally allow for detectable tensors if $n$ is sufficiently 
small, in practice this constraint would be further tightened by a more 
complete calculation. Nonetheless, our analysis does not necessarily 
rule out these models as viable candidates for inflation. Rather, it  
suggests that it will be difficult to construct a working model 
that results in a detectable tensor signal.

\section{Discussion}

The relativistic DBI brane scenario represents an attractive, 
string-inspired realisation of the inflationary scenario. Recent
cosmological data has placed very strong constraints on the simplest 
models based on a single ${\rm D3}$-brane. The strength 
of these constraints follows from field-theoretic upper limits 
on the tensor-scalar ratio, $r$, which in turn arise because 
the effective DBI action satisfies special  
algebraic properties. This provides motivation 
for considering generalisations of the scenario, in particular to 
multi-brane configurations. 

In this paper we have identified a phenomenological class of 
effective actions for which the constraints 
on $r$ are relaxed if significant (and detectable) 
non-Gaussian curvature perturbations are generated during inflation. 
Included in this class is the relativistic limit of the 
action associated with $n$ 
coincident branes in the small $n$ limit. Moreover 
we have found that such an effective action for arbitrary, finite $n$ 
can be expressed directly in terms of the corresponding action 
for the $n = 2$ model, due to the fact that the spin-$1/2$ representation
of $SU(2)$ is actually the fundamental one. This allows us to construct models for various values of $n$ using the
two-brane action and the iteration equations.
Physically these brane configurations typically 
have a smaller sound speed than the single brane models due to
the different structure of their action. This differing structure is also
manifest in the non-relativistic limit - since the non-Abelian nature of the theory
introduces new 'potential terms' that couple to the usual kinetic components of the 
action. In some cases this extra potential could help to further flatten the inflaton potential, 
whereas in other cases it will make it significantly steeper. An indepth analysis of slow roll in
such models would be welcome.
Their backreaction is also significantly smaller than other multi-brane 
configurations, and therefore this relaxes the amount of tuning 
required for the background charge.

We then proceeded to consider the question of whether the upper limits on 
$r$ could be relaxed to such an extent 
that a background of primordial gravitational waves 
might be detectable in future CMB experiments. The vast majority of 
string-inspired inflationary models that have been proposed to date 
generate an unobservable tensor background. We 
found that a detectable signal is possible, in principle, 
for typical string-theoretic parameter values 
if the number of coincident branes, $n$, is either $2$ or $3$. 
This is consistent with known F-theory configurations and 
current WMAP3 limits on the non-Gaussianity. Furthermore, 
we found that the level of non-Gaussianity must exceed $\fnl 
\lsim -70$ if such configurations are to generate a detectable tensor 
signal. This is well within the projected sensitivity 
of the Planck satellite.   

Our analysis invoked an $AdS_5 \times X_5$ warped throat geometry. However we 
made no assumptions regarding the form of the inflaton potential, other 
than imposing the implicit requirement that the universe underwent a phase of 
quasi-exponential expansion. In this sense, therefore, we have 
yet to explicitly establish that these inflationary models will 
be able to generate a measurable tensor signal. 
Nonetheless, since such a detection would provide a unique observational 
window into high energy physics, our results 
provide strong motivation for considering the cosmological consequences 
of these multi-brane configurations further when specific choices for the 
inflaton potential are made. In particular, it would be interesting 
to employ the techniques developed in
\cite{bean,Peiris:2007gz,Lorenz:2007ze,Bean:2007eh,Bean:2008ga} 
to identify the ranges of parameter space that are consistent 
with current cosmological observations. It would also be interesting to 
investigate whether the effective action (\ref{gensol}) with values of 
$m \ne - 1/2$ arises in other string-inspired settings. 

\section*{Acknowledgements}
We would like to thank Adam Ritz for useful comments on the manuscript.
IH is supported jointly by a Queen Mary studentship and the Science and
Technology Facilities Council (STFC).

\section*{Appendix: Exact Solution}

Eq. (\ref{defalpha}) can be analytically solved in full 
generality without imposing the limits (\ref{Plimits}) on the 
derivatives of the kinetic function. This allows us to determine the 
most general class of models where the non-linearity parameter 
satisfies the condition $\fnl \propto 1/c_s^2$ at leading order. 

In general Eq. (\ref{defalpha}) takes the form 
\begin{equation}
\label{gengen}
(2-\alpha ) P_{,X}P_{,XX} + 4XP^2_{,XX} = \frac{2\alpha }{3}
X P_{,X}P_{,XXX}
\end{equation}
and this reduces to 
\begin{equation}
\label{generalreduce}
\alpha Q_{,X} = (6-\alpha ) Q^2 + \frac{3(2-\alpha )}{2} \frac{Q}{X} \, ,
\end{equation}
where $Q \equiv P_{,XX}/P_{,X}$. 
Eq. (\ref{generalreduce}) can be transformed into the 
linear equation
\begin{equation}
\label{lineargeneral}
U_{,X}+ \frac{3(2-\alpha )}{2\alpha} \frac{U}{X} = \frac{\alpha -6}{\alpha}
\end{equation}
after the change of variables $U \equiv 1/Q$
and the general solution to Eq. (\ref{lineargeneral}) is given by 
\begin{equation}
\label{gensollinear}
\frac{P_{,XX}}{P_{,X}} = \frac{1}{X\left[ f_2(\phi) X^{(\alpha -6)/2\alpha}
-2 \right] } \, .
\end{equation}
Integrating a second time implies that
\begin{equation}
\label{secondint}
P_{,X} = f_1 (\phi ) \left( 1- f_2(\phi ) X^{-s} \right)^{1/(2s)}  \, ,
\end{equation}
where $s \equiv (\alpha -6 )/(2 \alpha)$ and we have redefined 
the arbitrary integration functions $f_i(\phi )$.  
Finally Eq. (\ref{secondint}) can be formally integrated 
in terms of a hypergeometric function 
\begin{equation}
 \label{thirdint}
 P= f_1X \,{_2}F_1 \left( -\frac{1}{s}, -\frac{1}{2s}; 1-\frac{1}{s}, f_2X^{-s}
\right)  \, ,
\end{equation} 
which represents the most general solution for this class of models. 
Note that we have set the
remaining constant of integration to zero to ensure 
that the kinetic function vanishes in the limit of
zero velocity. In fact this expression admits many 
different classes of solution, arising as limits
of the expansion of the hypergeometric function.

The special case of $\alpha =2$ $(s=-1)$ implies (after a 
further redefinition of the functions $f_i (\phi ))$ that 
\begin{equation}
\label{DBIsolution}
P = f_1 \sqrt{1-f_2 X} -f_3
\end{equation}
and this corresponds to the standard DBI action (\ref{DBIkinetic}) 
\cite{chenetal,lidser2}. 	  

The case $\alpha =18/5$ $(s=-1/3)$ can also be expressed in terms 
of elementary functions, again after redefinition of the $f_i (\phi)$: 
\begin{equation}
P = \frac{f_1\left[8 - 4f_2X^{1/3}
-\left(f_2X^{1/3}\right)^2\right]}{\sqrt{1-f_2X^{1/3}}} -f_3 \, .
\end{equation}
Note that this expression appears in a slightly different 
form to that in (\ref{unifiedgenP}). 
However in deriving (\ref{unifiedgenP}) we assumed the
relativistic limit, which in turn imposes a non-trivial 
relation between $X$ and $\phi$. Using this, and with a 
suitable redefinition of the functions, we can
easily transform the above expression into the required form.

\section*{References}

\providecommand{\newblock}{}

\end{document}